# INVESTIGATING CYBERSECURITY ISSUES IN ACTIVE TRAFFIC MANAGEMENT SYSTEMS

**Zulqarnain H. Khattak, MSCE.**
Doctoral Student/ Research Assistant
Center for Transportation Studies,
Department of Civil and Environmental Engineering
Thornton Hall D101, 351 McCormick Road
University of Virginia, Charlottesville, VA 22904
zk6cq@virginia.edu

**Hyungjun Park, Ph.D.**
Senior Scientist
Center for Transportation Studies,
Department of Civil and Environmental Engineering
Thornton Hall D101, 351 McCormick Road
University of Virginia, Charlottesville, VA 22904
hpark@email.virginia.edu

**Seongah Hong**
Doctoral Student/ Research Assistant
Center for Transportation Studies,
Department of Civil and Environmental Engineering
Thornton Hall D101, 351 McCormick Road
University of Virginia, Charlottesville, VA 22904
sh3zm@virginia.edu

**Richard Atta Boateng, MSCE.**
Doctoral Student/ Research Assistant
Center for Transportation Studies,
Department of Civil and Environmental Engineering
Thornton Hall D101, 351 McCormick Road
University of Virginia, Charlottesville, VA 22904
ra3fb@virginia.edu

**Brian L. Smith, Ph.D., P.E.**
Professor and Chair
Department of Civil and Environmental Engineering
Thornton Hall D101, 351 McCormick Road
University of Virginia, Charlottesville, VA 22904
briansmith@virginia.edu

**Word Count=5353, Figures and Tables=8*250=2000, Total = 7353**

## ABSTRACT

Active Traffic Management (ATM) systems have been introduced by transportation agencies to manage recurrent and non-recurrent congestion. ATM systems rely on the interconnectivity of components made possible by wired and/or wireless networks. Unfortunately, this connectivity that supports ATM systems also provides potential system access points that results in vulnerability to cyberattacks. This is becoming more pronounced as ATM systems begin to integrate internet of things (IoT) devices. Hence, there is a need to rigorously evaluate ATM systems for cyberattack vulnerabilities, and explore design concepts that provide stability and graceful degradation in the face of cyberattacks.

In this research, a prototype ATM system along with a real-time cyberattack monitoring system were developed for a 1.5-mile section of I-66 in Northern Virginia. The monitoring system detects deviation from expected operation of an ATM system by comparing lane control states generated by the ATM system with lane control states deemed most likely by the monitoring system. This comparison provides the functionality to continuously monitor the system for abnormalities that would result from a cyberattack. In case of any deviation between two sets of states, the monitoring system displays the lane control states generated by the back-up data source.

In a simulation experiment, the prototype ATM system and cyberattack monitoring system were subject to emulated cyberattacks. The evaluation results showed that the ATM system, when operating properly in the absence of attacks, improved average vehicle speed in the system to 60mph (a 13% increase compared to the baseline case without ATM). However, when subject to cyberattack, the mean speed reduced by 15% compared to the case with the ATM system and was similar to the baseline case. This illustrates that the effectiveness of the ATM system was negated by cyberattacks. The monitoring system however, allowed the ATM system to revert to an expected state with a mean speed of 59mph and reduced the negative impact of cyberattacks. These results illustrate the need to revisit ATM system design concepts as a means to protect against cyberattacks in addition to traditional system intrusion prevention approaches.

# INTRODUCTION

Cybersecurity is a broad term referring to processes and practices designed to protect networks, computers, programs and data from attack, damage or unauthorized access (*1*). In today's connected world, cybersecurity is an emerging threat in every field that relies upon communications. Therefore, transportation operation and management systems utilizing wired and wireless communications for managing roadways are also at significant risk of such cyberattacks. While the use of smart devices and communications provides new features and functionality, they also introduce new risks.

Arabo (*2*) stated that with the emergence of internet of things (IoT) devices, many of which are beginning to be used in traffic management systems, the traditional strategy of isolating control systems from other networks will no longer be feasible. Similarly, Jang-Jaccard and Nepal (*3*) also stated that the exponential growth of internet connections has led to a significant growth of cyberattacks. Traditionally malware attacks (which are secretly acting against the will of users) used to happen at a single point of surface amongst vulnerable components. Hence, rather than protecting each asset, the perimeter defense strategy (using firewall or antivirus) was utilized but such efforts are now ineffective. The major drawbacks of greater use of information and communications technology (ICT) are greater cybersecurity risks (*4*).

Transportation operations and management systems in the past were closed proprietary systems (isolated systems), having very limited cyber vulnerabilities. With the emergence of cloud or network computing and continued reliance on emerging technology such as IoT, those proprietary systems have now transformed into more open systems (*5*) with increased accessibility. Furthermore, the National Transportation Communication for ITS Protocol (NTCIP) uses center to center communications that rely on request based protocols through XML messages (*6*). These protocols have no security built into them and rely on the assumption that most attacks are from the inside and hackers make up only a small portion of total intrusions, thus, further increasing the risk over proprietary systems (*7*). In this regard, the USDOT has taken a huge initiative to develop security credential management system (SCMS) (*8*)- a message security solution for vehicle to vehicle (V2V) and vehicle to infrastructure (V2I) communication.

Active traffic management (ATM) involves the ability to manage recurrent and non-recurrent congestion using real time and predictive operational strategies. An ATM system can be deployed either in isolation to address a specific need such as the use of adaptive ramp metering to control the flow of traffic entering the freeway or can be used in conjunction with other systems to meet network wide needs of congestion management, incident management, traveler information and safety resulting in synergetic gains. Motorists rely on information received via ATM systems for making travel decisions. Thus, making the validity of such information even more critical but these modern open protocol systems leverages on continuous communication between roadway and traffic management center (TMC). This dependency on communication opens up a wide array of entry points, which makes them vulnerable to cyberattacks and are the least understood in terms of cybersecurity.

The research presented in this paper is based on the premise that perfect protection from cyberattacks is not realistic. Furthermore, a system may also be attacked internally by a disgruntled employee. Therefore, this research chose to develop a real-time threat monitoring system to continuously check to see if the system is behaving as expected. This paper is further structured as follows. First, a literature review is presented to analyze relevant studies and how different fields perceive cybersecurity. Then, the next section describes the effort made to develop a prototype ATM system. This is followed by a section that identifies various attack points in the prototype ATM system and presents examples of architecture for the monitoring system. Then next section explains the development of a monitoring system and finally an evaluation is conducted for identifying the impact of a possible attack on the prototype ATM system, and benefits of reverting back to a safe lane control state using a monitoring system.

## LITERATURE REVIEW

Although different fields have realized the importance of Cybersecurity and various studies have been conducted on cyber issues of control systems but transportation industry is still lagging in this field. Cherdantseva et al. (*9*) reviewed the state of the art practice in cybersecurity risk assessment of Supervisory Control and Data Acquisition (SCADA) systems and proposed that security was not the primary concern in standalone SCADA systems in the past, which was achieved by controlling physical access to system components. However, the situation has now changed and several standards and directives relating to the cybersecurity of SCADA systems have emerged. They examined the advantages and drawbacks of the analyzed risk assessment methods and concluded by stating that despite large number of methods for risk assessment of SCADA systems, there is still room for future research and improvement in terms of overcoming failure and developing monitoring systems. Similarly, Woo and Kim (*10*) provided quantitative assessment of cybersecurity risk of SCADA system based on optimal power flow in smart grid system and quantified vulnerabilities by first defining relevance of threat to each component and then assigning vulnerability index to each component based on historical data. The study finally calculated risk based on monetary terms as a product of probabilities of a threat and vulnerabilities, and the cost of an asset and concluded that this was a starting stage for quantifying cyber risks. C'ardenas et al. (*11*) used a chemical reactor for developing an anomaly detection monitoring system based on cyberattacks that change the behavior of targeted control system. The paper used simulation to test attacks and their impact on Tennessee Eastman process control system model using irreversible reaction inside a reactor of fixed volume and proposed an automatic response mechanism- that monitors the system. Another study evaluated the disruption of electric power system caused by cyberattacks while stating that cyber systems forms the backbone of nation's critical infrastructure. The paper stated that interdependence among computers, communication and power infrastructure have increased the risk of cyberattacks and proposed a real-time monitoring and anomaly detection system along with impact analysis. The impact analysis involved analysis of intrusion behaviors and evaluation of consequences of cyberattacks based on potential loss of power (*12*). Wu et al. (*13*) also used an integrated adaptive cybersecurity monitoring system to provide cyberattack awareness for scientific and military applications. The paper further used sensor data from networks and large number of simulations to evaluate the performance of the proposed monitoring system.

Similarly, Ashok et al. (*14*) proposed applications based on wide area monitoring system for power grid security. The paper provides architectural concepts pertaining to monitoring systems for anomaly detections and mitigation.

In recent years, several attacks have been reported on transportation operation and management systems compromising the safety (*15*) and security of these systems. During summer of 2011, a cyberattack carried out by a group known as "Anonymous" against a metropolitan transit system compromised personal identifiable information (PII) such as customer names, addresses, emails, ID numbers and passwords for more than 200 customers (*16*). During July 2009, several hardware experts revealed the vulnerabilities of a parking meter by hacking a parking meter in San Francisco with unlimited credit parking card using reverse engineering techniques (*17*). More recently, a group of experts demonstrated their ability to hack Jeep Cherokee connected car equipped with Uconnect system. Using the IP address, the two hackers were able to turn the engine of the car off, activate and de-activate the brakes, take remote control of the vehicle's information display and entertainment system, and activate the windshield wipers (*18*). Hacking of dynamic message signs (DMS) are frequently in the news, and on recent Memorial Day Weekend, again DMS signs were hacked with signs ranging from political slogans to "Party Yardy Hall!" and a week later displaying a statement regarding recent Cincinnati Zoo gorilla incident (*19*).

With these attacks on transportation infrastructure systems over the past several years, the conventional wisdom is that cybersecurity is a critical and growing issue. Yet, there has been little research in the area. Folk (*20*) first provided a brief overview of cybersecurity issues in transportation systems along with a history of some past attacks and later (*21*) provided fundamentals of common cyberattack process and offered ideas on how to prepare and respond to an incident. Similarly, Amoozadeh et al. (*22*) evaluated the effects of cyberattacks on communication channel as well as sensor tampering of a connected vehicle (CV) stream equipped to achieve cooperative adaptive cruise control (CACC) using a simulation model. Their results revealed that an insider attack can cause significant instability in CACC vehicle stream and they also proposed various countermeasures such as downgrading to adaptive cruise control (ACC), that could be used to improve security and safety of connected vehicles (*23*). Since there is no way to eliminate cyberattacks completely and no rigorous study exist on cyber vulnerabilities of transportation operations and management systems, this paper tries to fill this gap in literature by providing a first attempt to evaluate the impact of cyber vulnerabilities of transportation operation and management systems and proposing a real-time monitoring system for detecting cyberattacks to avoid the consequences of a compromised system. It should be noted that deploying monitoring system is a well-accepted concept in cybersecurity risk assessment for detecting anomalies (*12–14*).

## STUDY LOCATION

The ATM system deployed on I-66 from Washington, D.C line to Route 29 in Gainesville, Virginia was used as a case study in this research and is presented in Figure 1. The ATM system consists of speed harmonization, queue warning, hard shoulder running and lane control states. These systems were deployed for improving safety and incident management at heavy bottleneck locations that averaged 60,000 vehicles per day between US-15 and Fairfax Line. The number further increased to 90,000 vehicles per day between the Fairfax line and I-495 while reduced to 64,500 vehicles per day between I-495 and Arlington Co-Line. Similarly, highest crash frequency existed at SR 7100 and I-495. This study modeled a 1.5-mile section of the I-66 ATM system in VISSIM for cybersecurity evaluation.

## DEVELOPMENT OF A PROTOTYPE ATM SYSTEM

A prototype ATM system was developed first based on the concept of operations of the actual ATM system on I-66. The various steps involved are presented in Figure 2 and a detailed description for each step is provided below.

### VISSIM Simulation

A microscopic traffic simulation of the corridor was developed to serve as the "real world" in our prototype system. This allowed the research team to see the impact of the ATM system, both without attacks and under attack, on the transportation system. This step involved the modeling of test bed network in microsimulation software VISSIM shown in Figure 3. The testbed network consists of a 1.5-mile segment of a freeway corridor with three lanes in one direction. For the variable lane control state system, two gantries (each with a set of three states displaying either open, closed, or merge) were placed at the mileposts 0.5mile and 1mile along the corridor. A traffic incident was modeled within the most downstream section between milepost 1 and 1.5 miles. The network was loaded with a vehicular volume of 3000, 4500 and 6000 vehicles per hour to represent low, medium and high congestion conditions of traffic.

### Data Collection

During this step, various detector stations were embedded in the VISSIM network at each lane near each gantry. The detectors recorded traffic measurements such as speed and traffic volume, which were aggregated within the predetermined data collection interval (every 10 sec). Using the COM interface, the traffic measurement data was retrieved at every update instance and sent to the ATM algorithm.

### Data Analysis

In this step, the data collected from the simulation was processed and analyzed to identify whether there is a disruption of flow on the network. The analysis was based on speed of vehicles dropping below a threshold or speed at one instant of time being lower than the speed at preceding instant of time. The algorithm used for incident detection is provided below;

$$U_{i,t} == 0$$
$$Q_{i,t} == 0 \; or,$$
$$U_{i,t} < C_1 * U_{i,t-1} \text{ and } U_{j,t} < \pm(1 + C_2) * U_{j,t-1}$$

Where U denotes speed, Q denotes volume, i represents lane under study and j denotes the other two lanes. Similarly, t represents a particular instant of time and t-1 represents the immediately preceding instant of time. Hence, the algorithm detects incidents if speeds or volume on a lane under consideration drops to zero or the speed at one particular instant of time is lower than another instant of time and the speed in the opposing two lanes at that instant of time is greater than the speed at the preceding instant of time.

**Selection of Lane Control States Using Decision Tree**

During this step, the lane control states were selected based on the set of decision trees developed based on the concept of operations of the actual ATM system on I-66 – see Figure 4. The lane control policy was decided based on the traffic speed collected through detector stations embedded in the VISSIM model while the variable lane control state application was modeled using VISSIM API and C# language. This policy is consistent with ATM control used for operation on I-66 (*24*).  The decision tree used for selecting lane control states consist of four critical decision variables namely, speed, event, distance to the event and the lane on which the event is occurring. The first critical variable to be screened for deciding the appropriate lane control state for display is speed and if for instance, a speed greater than 20mph is selected, the next critical variable of event is screened. It has three decision variables namely, no event, paving work on the road, and an incident on the road. After selecting the type of event, the third critical decision variable of distance to the event is screened by the algorithm. If for instance, the event is less than 0.2 miles away from the gantry location, the last decision variable is screened to find the lane on which the event is occurring. Once these critical variables are screened and appropriate decision variables are selected, the applicable lane control states for display are selected for each of the three lanes.

**Implementation of Selected States Back to the Simulation**

Once the states for lane control were selected using the data from simulation model and its screening for appropriate traffic conditions based on the decision tree, they were transferred back to the VISSIM network for developing a feedback loop that operates in continuity. The driving behavior of the vehicles was also controlled according to the selected lane control states using VISSIM API. The vehicles at the upstream location, which were required to comply with the variable lane control states were controlled to conduct lane merge to the adjacent lane when it was recommended by the decision of the lane control states application and were not allowed to travel in specific lanes when the lane control state algorithm indicated a lane closure.

## DEVELOPMENT OF A MONITORING SYSTEM

The objective of this step was to develop a monitoring system, starting with the identification of entry points for cyberattacks within the ATM system, and then ending with proposing examples for the development of monitoring system architecture based on the identified attack points. Based on the results of this step, a prototype monitoring system was finally developed.

**Identification of Entry Points**

The identification of entry points requires clear description of the ATM architecture in terms of its data flow. The data flow diagram in Figure 5 shows the functional aspects of lane control states along with likely entry points for cyberattacks. The lane control states are connected to the network switches in the cabinets provided at each of the gantry locations, from where they are again connected into the existing fiber optic cable that runs to the TMC where the ATM software is installed. The ATM system is built around data collection from various roadway traffic detectors, environmental condition detectors, and monitoring through CCTV. The data collected is used to detect any unusual behavior such as congestion and incidents, which is processed by the ATM software using predefined logic to come up with appropriate response plans in terms of which lane control states to display on the road network. Additional connections to private sector probe data providers, such as Inrix, and Internet of things (IoT) devices also exist. The testbed network developed in VISSIM mimics the data flow architecture of I-66 ATM system, with vehicle detectors as the main data source. The attack points that are likely to be compromised by an adversary in the prototype system includes point A (the ATM software itself), point B (data source such as detectors), and point C (connection between field device, lane control states and cabinet at gantry location). The adversary can affect the normal operation of the lane control states by gaining access to any of these three attack points for instance, gaining access to the ATM software through point A can allow the adversary to change the final decision generated by the ATM software algorithm and send any desired set of states for display on the gantries. Similarly, gaining access to the detectors through point B can allow adversary to send manipulated set of traffic data to the ATM software that may lead to generation of wrong states for display. Finally, gaining access to the gantry display through point C may allow the adversary to manipulate the decision for display of lane control states received from the ATM software algorithm and display any desired states on the gantries.

**Development of Monitoring System Architecture**

It is important to make it clear that the monitoring system developed in this research is tasked with monitoring the operation of the ATM system itself. We traditionally think of monitoring in traffic management as the use of detectors to measure speed and volume of vehicles. This provides an understanding of the system state. In a cybersecurity monitoring system – we are monitoring the functioning of the ATM system itself to be sure it is behaving as expected. Furthermore, the main purpose of developing a monitoring system is to add further redundancy to the existing system so as to make it difficult for an attacker to achieve his or her motives. The architecture of the proposed monitoring system for the identified attack points is shown in Figure 6. Figure 6a and 6b provide examples of separate monitoring systems for attack points A and B. The proposed monitoring systems consists of two channels that process the traffic data independently; (a) one channel is associated with the operations of ATM application and (b) another channel is associated with the monitoring system, which decides the control decision and display states independently and matches the decision with that of the other channel associated with the ATM application. The two channels are communication paths, receiving data from two separate or same sources that is further processed for deciding lane control states. The difference between the two-individual monitoring systems shown in Figure 6a and 6b is in the data source processed by the monitoring channel i.e. the monitoring system for attack point A has only one data source of detectors, which is fed into both monitoring and ATM channel while the monitoring system for attack point B uses a different data source such as (historical data, CV data or data from CCTV) for the monitoring channel.

Similarly, the example for architecture of monitoring system proposed for attack point C is provided in Figure 6c. The monitoring system will again operate on the same principle with two different channels processing data independently (a) one channel associated with the operation of ATM application and (b) another channel associated with the monitoring system, which decides the control decision and display lane control states independently and matches the decision with that of the other channel associated with the ATM application. Here the states being displayed on the gantry are viewed through CCTV and matched with the states generated by the channel associated with the monitoring system. Based on the matching step, the possibility of a cyberattack will be decided like the previous monitoring system and in case of a cyberattack, lane control states selected by the monitoring channel will be displayed on the network and the operator will also be alerted of a possible cyberattack.

**Development of a Prototype Monitoring System**

Based on the example architectures proposed in previous section, a prototype monitoring system was developed for the ATM system. Figure 7 represents the architecture of the prototype monitoring system for multiple attack points A and B developed in this project. The traffic data is processed by two different channels; (a) one channel is associated with the operations of ATM application that receives data from detectors embedded in the network and (b) another channel is associated with the monitoring system, which receives data from CVs. Both channels decide the control decision and display lane control states independently. The lane control states generated by the two independent channels are matched by the monitoring system for detection of anomalies. It should be noted that both the monitoring channel and the ATM channel operate on the same logic but utilize the data from different sources. A 100% CV penetration was assumed in the simulation environment and data from individual vehicles were sent to the monitoring channel for generating lane control states. If any type of deviation is observed during the matching step, the monitoring system report it as an anomaly, or a cyberattack is possibly occurring. In case of anomaly detection, the monitoring system cedes the states generated by ATM channel and displays the lane control states generated by monitoring channel over the network. The monitoring system also provides an alert to the operator but if no deviation is observed during the matching step, the lane control states generated by the ATM algorithm are displayed on the network.

## EVALUATION RESULTS

The impact of a possible attack on the prototype ATM system, reverting back to a safe lane control state using a monitoring system and running the ATM system without a monitoring system was evaluated using VISSIM microscopic simulations. In this evaluation, four cases were prepared and used, depending on the existence of an ATM system, cyberattacks, and a monitoring system. These cases were:

- Baseline: without an ATM system, cyberattacks, and a monitoring system.
- Case 1: with an ATM system; but without cyberattacks, and a monitoring system.
- Case 2: with an ATM system, and cyberattacks; but without a monitoring system.
- Case 3: with an ATM system, cyberattacks, and a monitoring system.

For all the cases, an incident was created at the far downstream location of the 1.5mile test freeway segment, in an attempt to create a situation where an ATM system is needed. The ATM system described in the previous section was modeled in VISSIM and cyberattacks were randomly generated during the simulation using the COM interface and C# in Visual Studio. For the attacks at point A, the lane control states were randomly changed while, for the attacks at point B, the detector data was randomly scrambled. Lastly, if any anomalies in lane control states were detected, the monitoring system reverted the compromised lane control states back to the ones created by the monitoring channel using a different source of data, i.e. CV data. 55 replications for each of the four cases, with each replication running for 10 minutes with a warm-up period of 2 minutes, were conducted and the results are provided below.

This evaluation basically involved the comparison of mean speeds from all four cases using paired t-tests, assuming a null hypothesis that the two means were equal. The results are presented in Table 1 and the interpretations are provided below:

- Baseline: without an ATM system, cyberattacks, and a monitoring system
  - o This is the base case without an ATM system. Hence no attacks can happen and thus a monitoring system is not needed.
  - o A mean speed of 53mph was obtained in this Baseline case.
- Case 1: with an ATM system; but without cyberattacks, and a monitoring system
  - o This Case 1 added an ATM system on top of the Baseline case. The ATM system provided a set of lane control states to better manage and operate traffic according to the prevailing traffic situations.

- o It was found out that, with an ATM system in operation, the mean speed has improved to 60mph (13% increase from 53 mph of Baseline). This is statically significant with a p-value less than 0.01.
- Case 2: with an ATM system, and cyberattacks; but without a monitoring system
  - o In Case 2, random cyberattacks were generated, which compromised the original ATM lane control states and resulted in a set of "random" lane control states that may have a negative impact on traffic.
  - o The mean speed in Case 2 under attacks was degraded to 51mph, a reduction of 15% from 60mph of Case 1. This degradation was found to be statistically significant (p-value <0.01).
  - o Also, the Case 2 mean speed (51mph) was similar to the Baseline mean speed (53mph), implying that the effectiveness of the ATM was negated by attacks.
- Case 3: with an ATM system, cyberattacks, and a monitoring system
  - o Lastly, Case 3 added a monitoring system on Case 2. The monitoring system detected anomalies in lane control states created by attacks, and rectified the compromised states as needed.
  - o With the monitoring system, the mean speed was improved back to 59mph, 16% increase compared to Case 2 with a statistical significance (p-value <0.01).
  - o This Case 3 mean speed of 59mph was also found to be similar to the Case 1 speed of 60mph. The implication is that the monitoring system effectively overcame the negative impact created by cyberattacks and allowed the ATM system to function as intended.

In summary, this research has demonstrated that a cyberattack monitoring system has the potential to limit the impact of attacks on an ATM system and improve freeway operations.

## CONCLUSIONS

The research effort presented in this paper is a first step of a huge undertaking; i.e. investigating, analyzing, and addressing cyber security issues expected in various transportation-related systems in the very near future, from a holistic point of view. The goal here was to demonstrate the issues and a possible solution to highlight the significance of the cyber security analysis, rather than solving the problems. For this, we took the ATM system as an example and the simulation results were presented for this purpose.

This study is one of the first attempts to investigate cybersecurity issues in Active Traffic Management systems. A 1.5-mile section of the ATM system deployed on I-66 in Northern Virginia was selected as a case study. First, a prototype ATM system was developed based on the concept of operations of the actual ATM system in Northern Virginia. Then a prototype monitoring system was developed to detect anomalies in ATM operations and to avoid the consequences of a compromised system by cyberattacks. The monitoring system leverages on real time data by comparing the lane control states generated by two different channels, i.e. the ATM channel and the monitoring channel. In case of any deviation between two sets of lane control states, the monitoring system starts displaying the lane control states generated by the CV data and alerts the operator.

The evaluation results showed that the ATM system could increase the mean vehicle speed in the system by 13% compared to the baseline case. However, when subject to cyberattack through manipulation of traffic data or the lane control states through the ATM software, mean speed was similar to the baseline case and reduced by 15% compared to the case with the ATM system. This illustrates that the effectiveness of the ATM system was negated by cyberattacks. The monitoring system however, allowed the ATM system to revert to an expected state with a mean speed of 59mph and reduced the negative impact of cyberattacks. It is therefore, reasonable to say that the proposed monitoring system was successful in avoiding the severe consequences of cyberattacks and has significant potential in improving freeway operations. Hence, it is necessary to further continue the effort for developing and evaluating more rigorous monitoring systems, along with revisiting ATM system design concepts as a means to protect against cyberattacks in addition to traditional system intrusion prevention approaches.

## Author Contribution

Zulqarnain contributed by providing conceptual design and methodological framework, helping with simulation and data analysis, writing the whole draft manuscript and doing subsequent revisions. Seongah helped with the simulation coding. Richard helped with analyzing the data that came out of simulation. Prof. Smith and Dr. Park supervised the project and provided research directions. All authors reviewed the results and approved the final version of the manuscript.

## List of Tables

## List of Figures

Khattak, Park, Hong, Boateng, and Smith

**Table 1 Evaluation Results for ATM Monitoring System**

| Scenario | Presence of | | | Mean Speed (mph) | Standard Deviation | % Changes p-value | % Changes p-value | % Changes p-value |
|---|---|---|---|---|---|---|---|---|
| | **ATM** | **Cyber-attack** | **Monitoring System** | | | | | |
| **Baseline** | No | No | No | 53 | 20 | Base value compared to | N/A | N/A |
| **Case 1** | Yes | No | No | 60 | 14 | 13% <0.01 | Base value compared to | N/A |
| **Case 2** | Yes | Yes | No | 51 | 24 | -4% >0.11 | -15% <0.01 | Base value compared to |
| **Case 3** | Yes | Yes | Yes | 59 | 15 | 11% <0.01 | -2% >0.11 | 16% <0.01 |

Khattak, Park, Hong, Boateng, and Smith

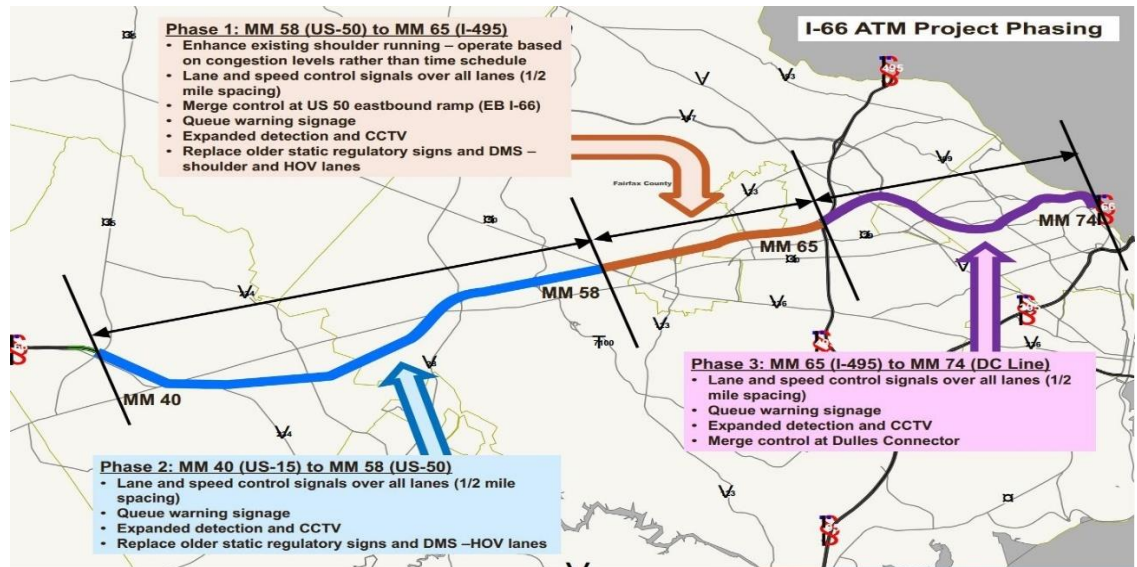

**FIGURE 1 I-66 Active Traffic Management System Location** (*7*)

Khattak, Park, Hong, Boateng, and Smith

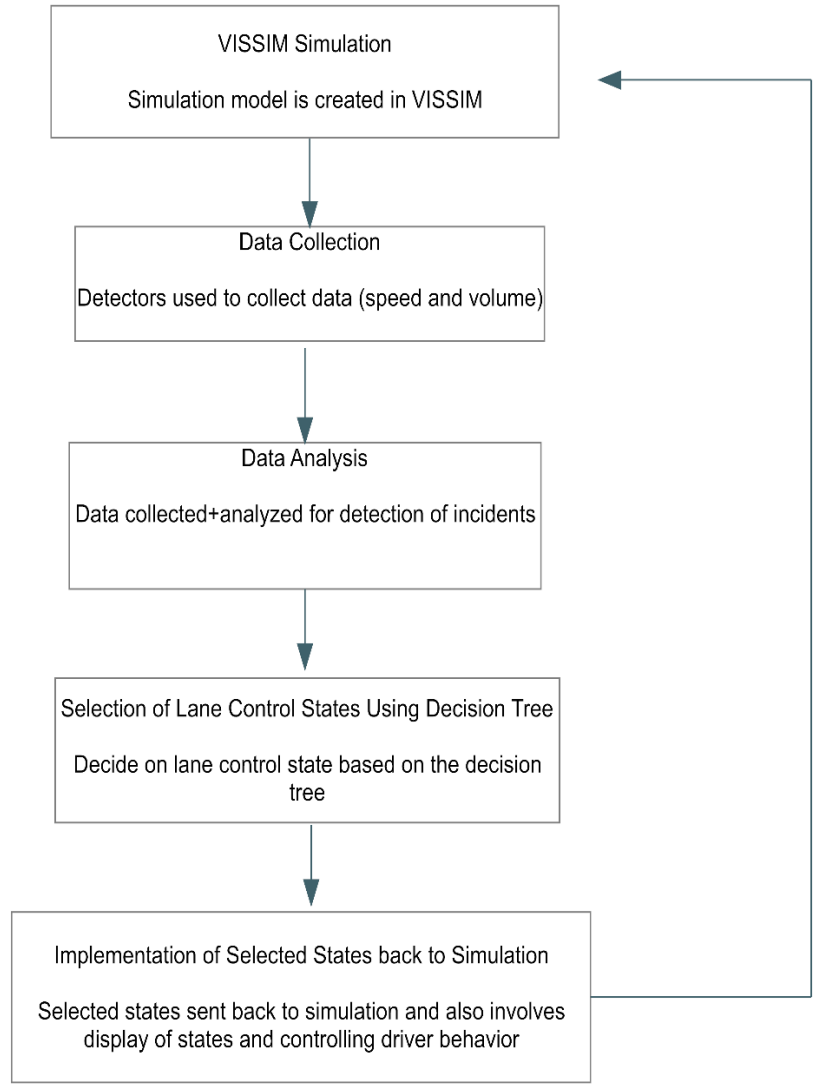

**FIGURE 2 Prototype ATM System Flow Chart**

Khattak, Park, Hong, Boateng, and Smith

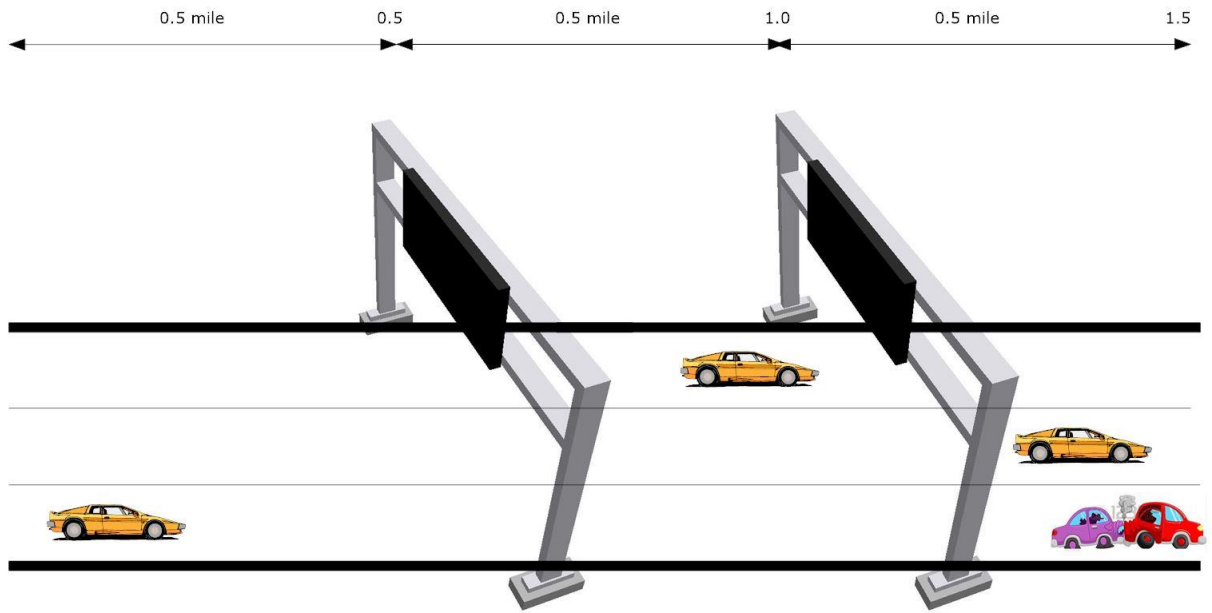

**FIGURE 3 Testbed Network**

Khattak, Park, Hong, Boateng, and Smith

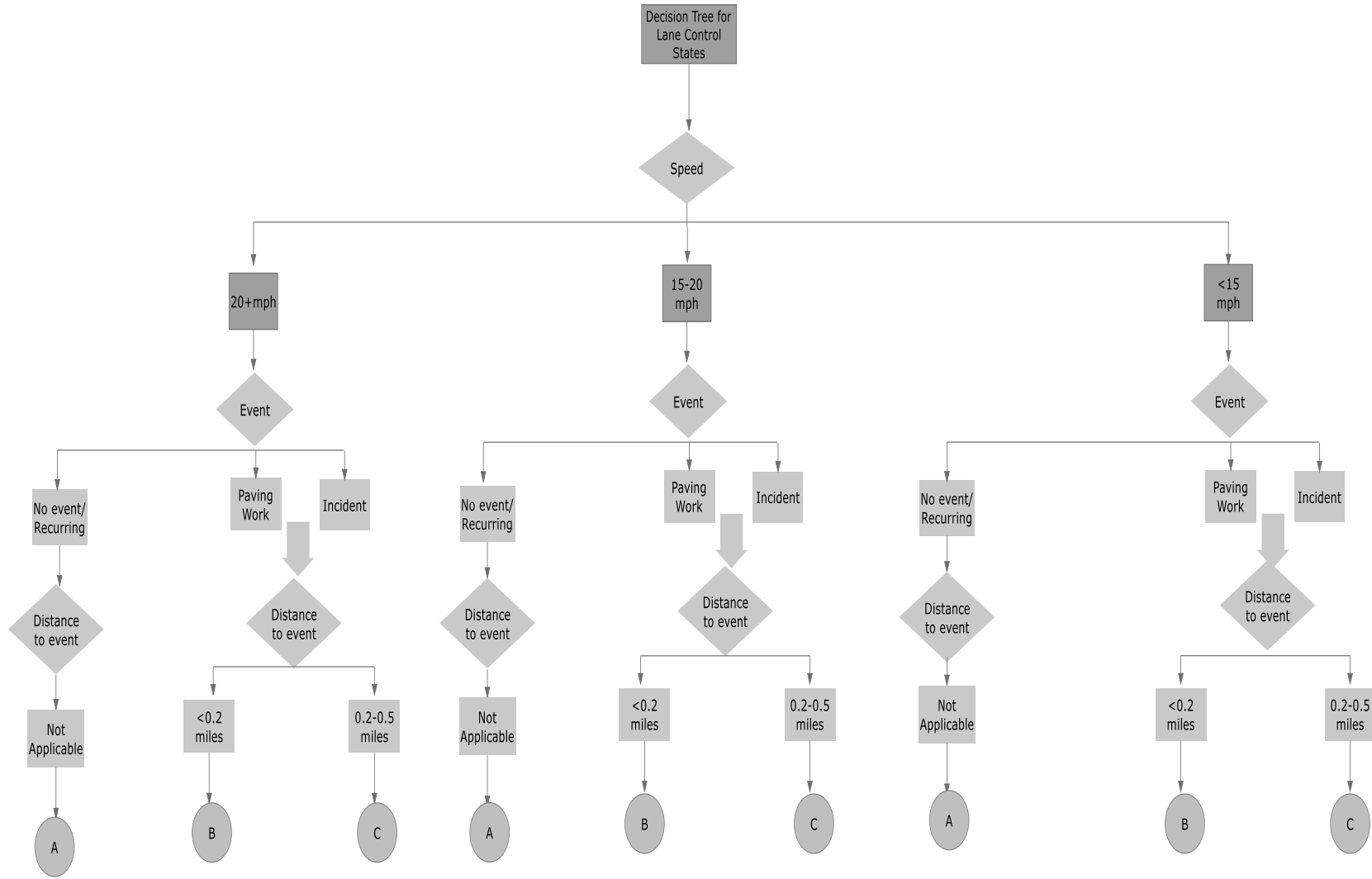

a) First three Decision Variables for Lane Control State
FIGURE 4 Decision Tree for Lane Control States

Khattak, Park, Hong, Boateng, and Smith

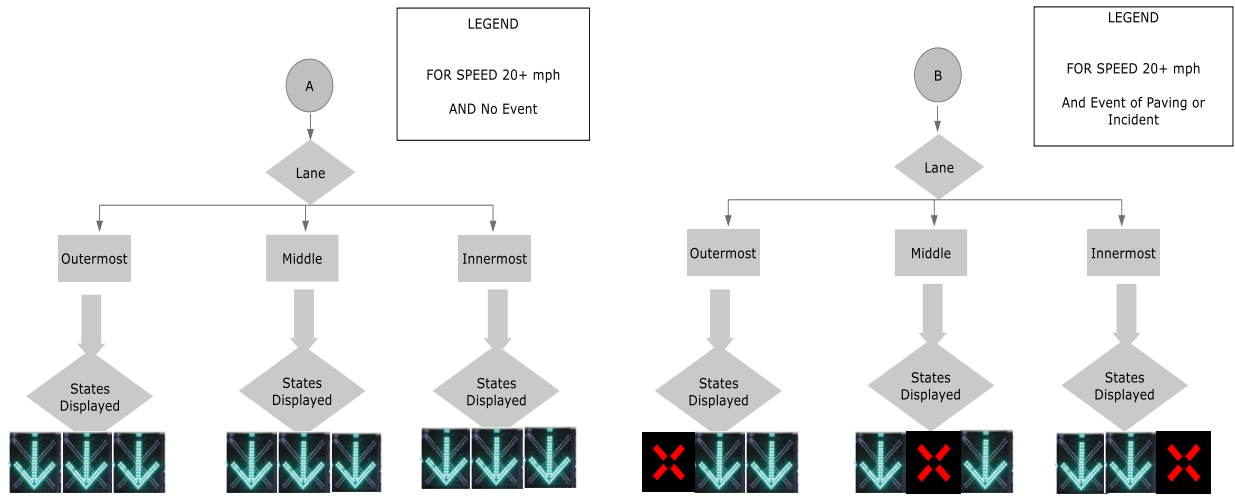

**b) An Example of Fourth Decision Variable for Speed of 20+ mph**
**FIGURE 4 Decision Tree for Lane Control States**

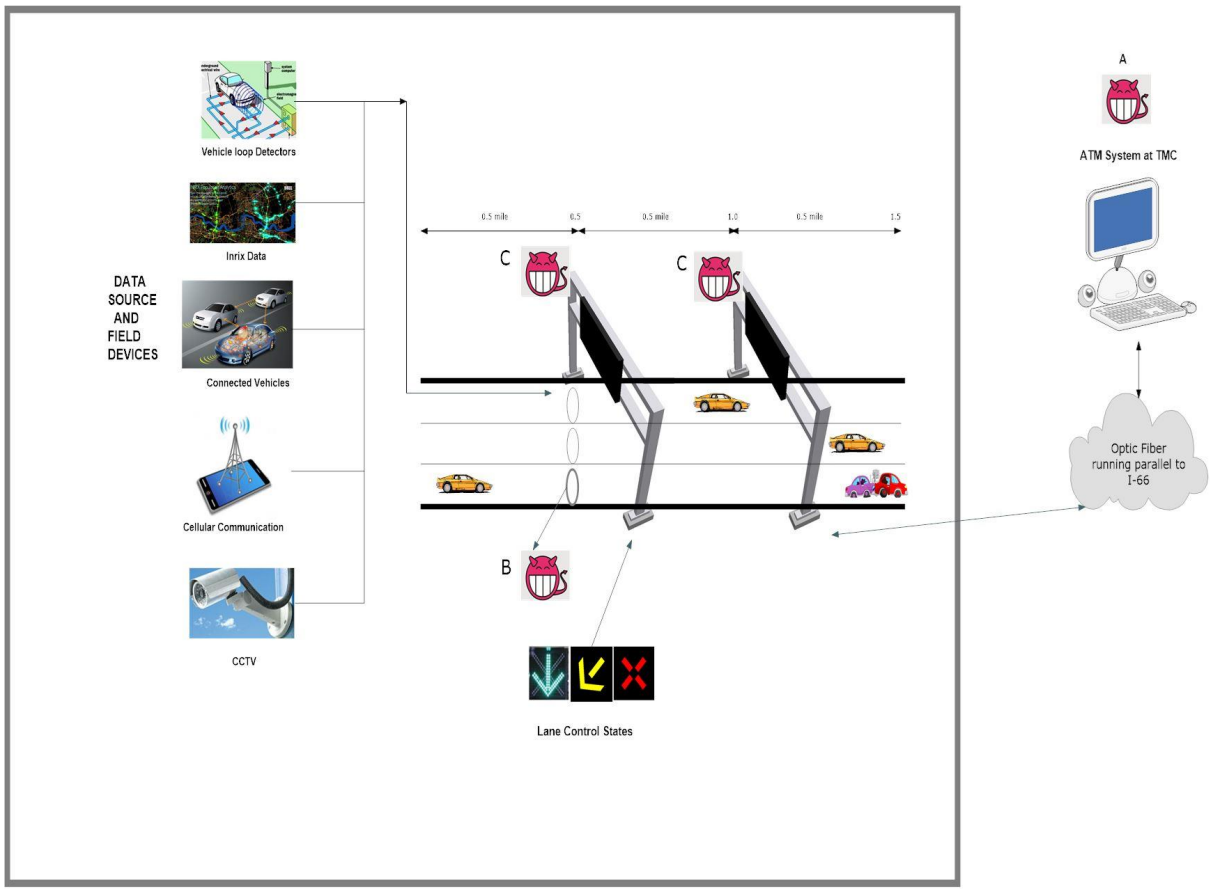

**FIGURE 5 Data Flow Diagram of ATM System with Attack Points**

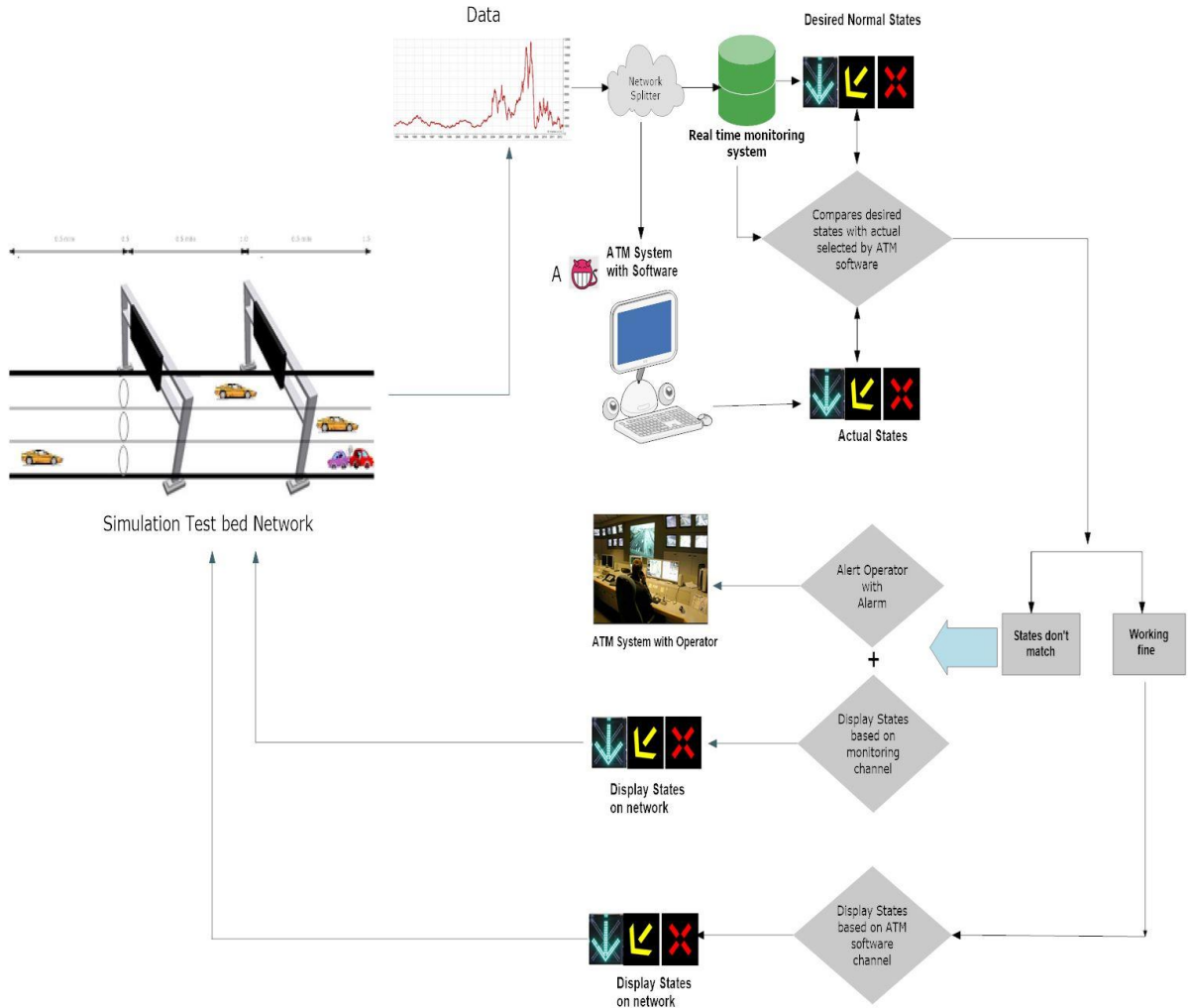

**a) Architecture of Monitoring System for Attack Point A (ATM Software at TMC)**
**FIGURE 6 Example Architecture of Monitoring Systems**

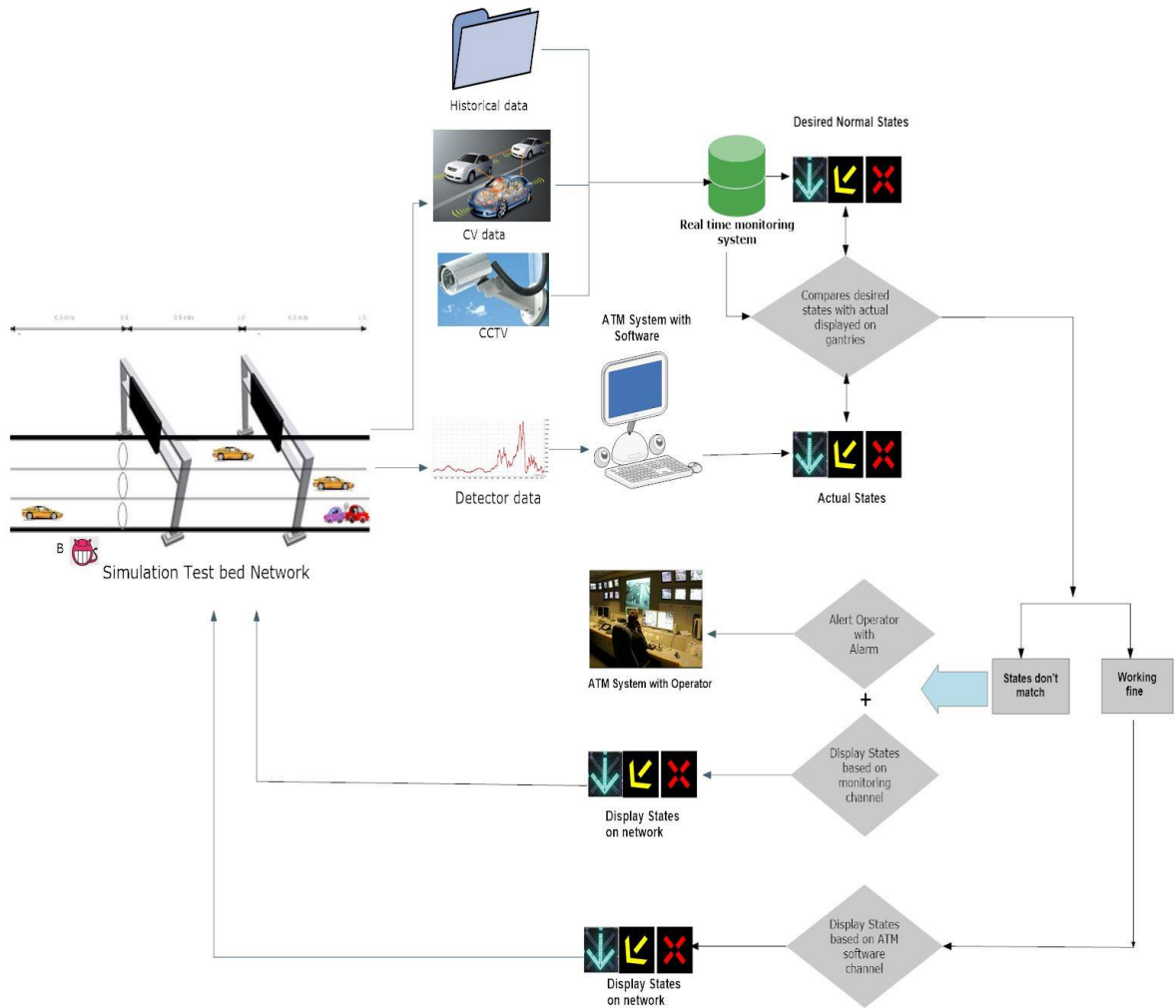

**b) Architecture of Monitoring System for Attack Point B (Detectors)**

**FIGURE 6 Example Architecture of Monitoring Systems**

Khattak, Park, Hong, Boateng, and Smith

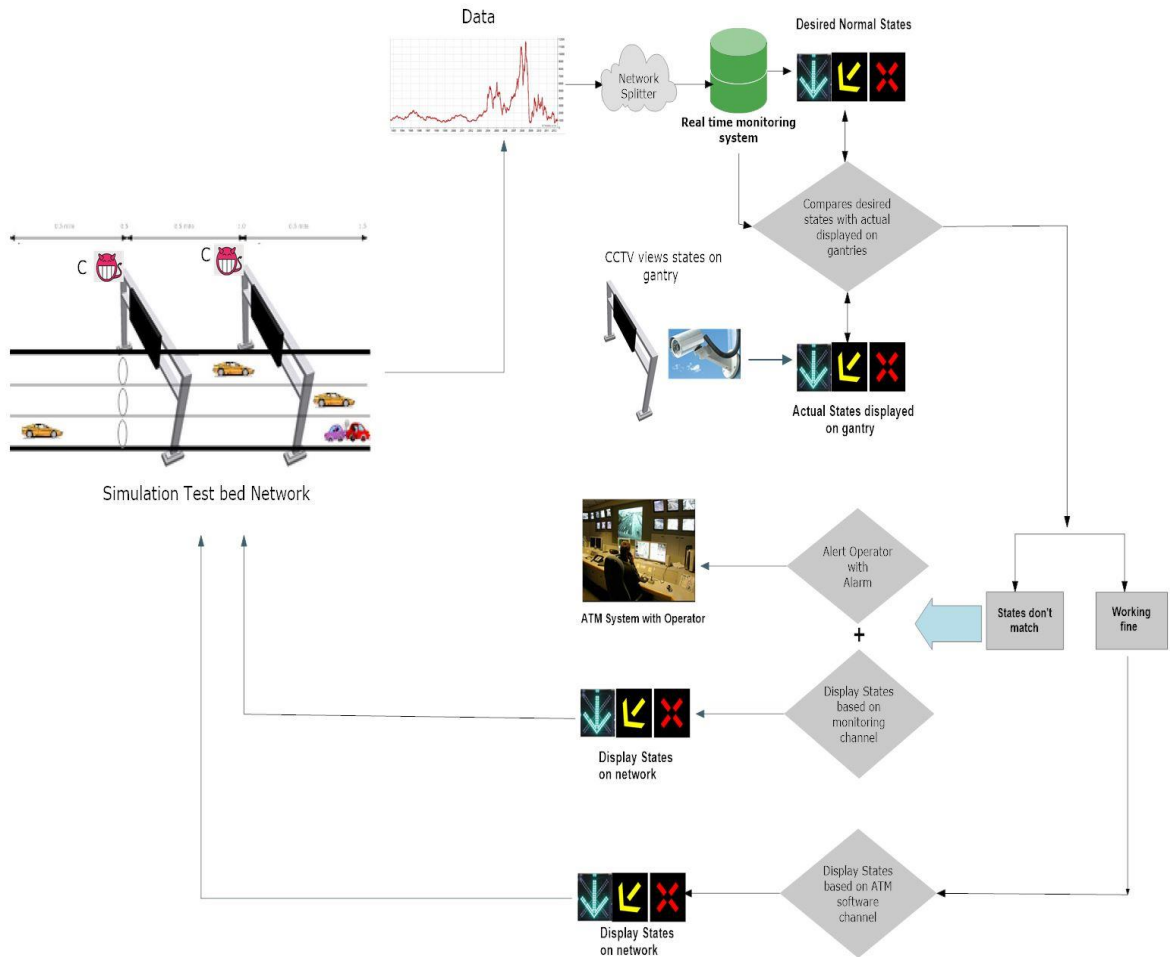

**c) Architecture of Monitoring System for Attack Point C (gantry locations)**

**FIGURE 6 Example Architecture of Monitoring Systems**

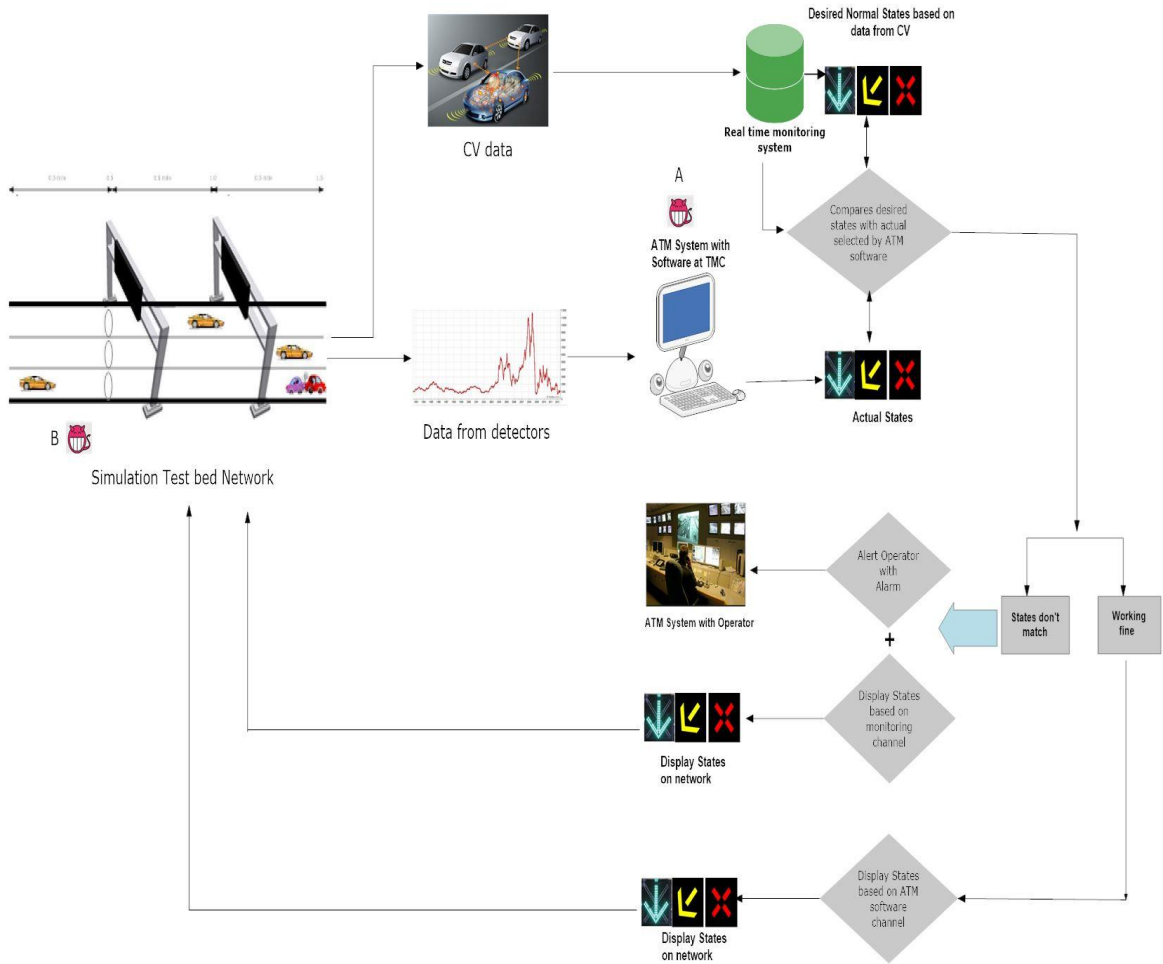

**FIGURE 7 Architecture of Monitoring System Covering Attack Points A and B**